\documentclass[aps,nofootinbib]{revtex4}

\usepackage{epsfig}
\usepackage{epsf}
\usepackage{bm}                 
\usepackage{amsmath}
\usepackage{amssymb}
\usepackage{dcolumn}
\usepackage{graphicx}
\usepackage{xfrac}
\usepackage{xcolor}
\usepackage[small,bf, format=plain, width=.75\textwidth, justification=RaggedRight]{caption}[2008/04/01]
\input amssym.def
\input amssym

\begin{document}

\title{Limits and Signatures of Relativistic Spaceflight}
\author{Ulvi Yurtsever} \email{ulvi.yurtsever@raytheon.com}
\author{Steven Wilkinson} \email{srwilkinson@raytheon.com}
\affiliation{\noindent
Raytheon Company, El Segundo, CA 92195}

\date{\today}

\begin{abstract}
While special relativity imposes an absolute speed limit at the speed of light, our Universe
is not empty Minkowski spacetime. The constituents that fill the interstellar/intergalactic
vacuum, including
the cosmic microwave background photons, impose a lower speed limit on any object travelling at
relativistic velocities. Scattering of cosmic microwave phtotons from an ultra-relativistic object
may create radiation with a characteristic signature allowing the detection of such objects
at large distances.

\end{abstract}


\maketitle
\thispagestyle{plain}

{\noindent \bf Introduction}

At a fixed speed $v$, the distance travelled by a spacecraft
within a given (proper) travel time $\tau$ scales as $\gamma v \tau$,
which for relativistic
speeds approximately equals $\gamma c \tau$, where
$\gamma=(1-v^2 /c^2 )^{-\sfrac{1}{2}}$. Hence range is directly
proportional to $\gamma$, and to cover intergalactic distances
within a limited ``lifetime" $\tau$ requires $\gamma \gg 1$.
The primary obstacles to relativistic
($\gamma \gg 1$) space travel would be collisions with
interstellar dust particles (cosmic dust) and larger space objects, which will impact with
kinetic energies
of $(\gamma -1) m c^2$ for rest mass $m$, similar to collisions in particle
accelerators but potentially at much higher energies.
For large enough $\gamma$,
even molecular collisions could be  a significant source of drag
and possibly damaging. For example,
at $\gamma = 2$ a baseball size object of mass 150g 
has an impact energy equivalent to 36 Megatons of TNT;
a single cosmic dust grain of mass $10^{-14}\rm g$ at
$\gamma = 10^{8}$ has an impact energy of
close to 24 kgs of TNT.~\cite{tiplerbook,weinbergbook,griffithsbook,perkinsbook}

\begin{figure}[htp]
\vspace*{-0.1in}
\hspace{1.6in}
\centerline{
\input epsf
\setlength{\epsfxsize}{7.425in}
\setlength{\epsfysize}{4.800in}
\epsffile{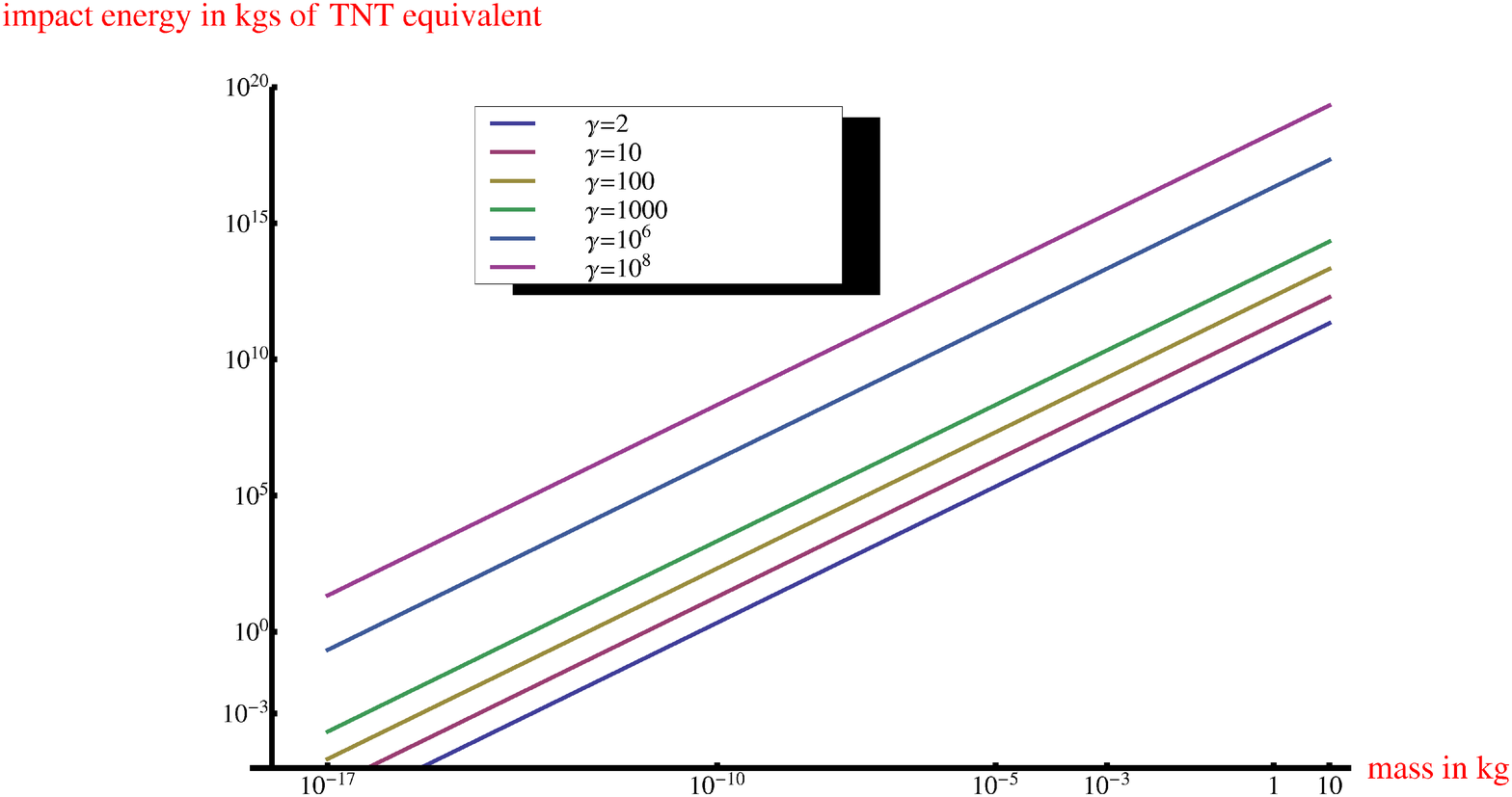}
}
\vspace{-1.0in}
\end{figure}

~

\begin{figure}[htp]
\vspace*{-0.6in}
\hspace{1.6in}
\centerline{
\input epsf
\setlength{\epsfxsize}{7.425in}
\setlength{\epsfysize}{4.800in}
\epsffile{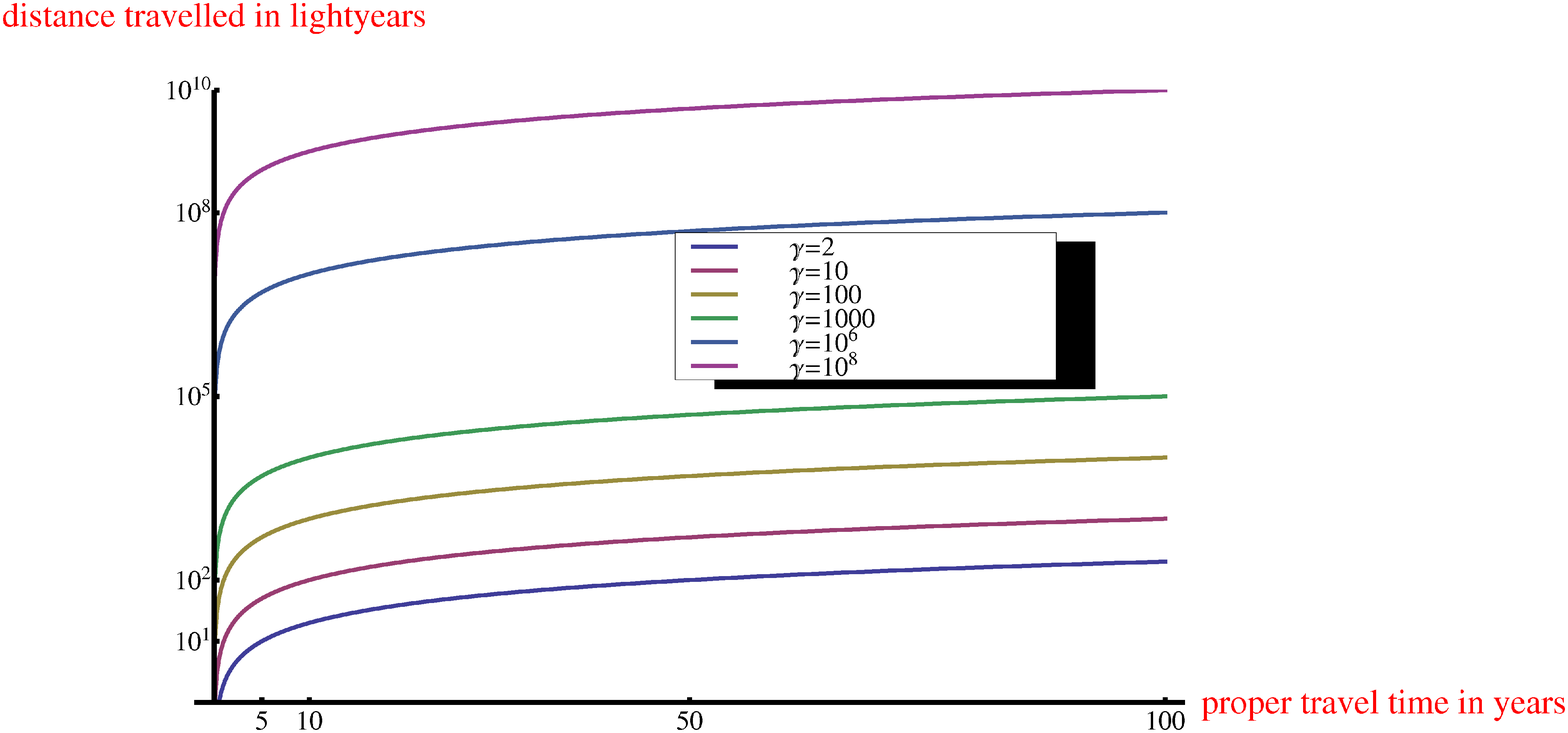}
}
\vspace{-0.9in}
\end{figure}
The intergalactic medium has much less debris per unit volume
as compared to the interstellar medium so the chance of a
catastrophic collision occurring is significantly reduced
beyond the confines of any galaxy.
We define extreme velocities to occur at $\gamma > 9.1$
which is the threshold velocity for proton-proton collisions to produce
antiprotons. Now to travel at a safe speed
in the interstellar medium one would need a $\gamma < 1.3$,
the threshold for the pion production in proton-proton collisions.
The problem of colliding with larger objects as described above remains,
but we assume that sufficiently advanced technology
likely to accompany relativistic space travel capabilitites will be able to circumvent
these interactions.
One can imagine, though, intergalactic travel through mostly
matter-free space, and under those conditions the
cosmic microwave background (CMB) photons
are primarily the particles a spacecraft will encounter in collisions.

~

{\noindent \bf CMB Interactions}

A fast-moving spacecraft traveling in intergalactic space still
has to contend with
collisions with cosmic microwave photons, which, at relativistic speeds,
will appear in the spacecraft frame as highly energetic gamma rays.
Interactions of CMB photons with the material of the spacecraft hull will have effects
ranging from ionization and Compton scattering to pair production with increasing $\gamma$.
We will assume that advanced technology can mitigate the harmful
effets of ionization and Compton scattering, and therefore concentrate on pair production
as the obstacle that is most likely to resist removal via technology.~\cite{tiplerbook}

The first speed level where pair production will pose
a challenge for relativistic spaceflight engineers
is photon-nucleon interactions
in the hull of the spacecraft.
Let us consider a hull made of ordinary baryonic
matter. In this case,
a single CMB photon will create an
electron-positron pair via its collision with a nucleus
in the hull if its
energy exceeds the rest mass of an electron and positron. That threshold is
at blueshifted photon energies of about 1 MeV.
In terms of the frequency $\omega$ in the rest frame of the spacecraft, the condition
for nucleus-mediated pair production is
\begin{equation}
\omega \geqslant \frac{2 m_e c^2}{\hbar}
\approx 2.47 \times 10^{20} \mbox{Hz} 
\end{equation}
The central frequency of the (Planck) distribution of
cosmic microwave background photons is 160GHz.
The blueshift factor to transform a typical counter-directed
CMB photon to the rest frame of the spacecraft
is then $\omega=\gamma (1+v/c) \times 10^{12}\, \mbox{Hz}$, where $v$ is the velocity of the
spacecraft and $\gamma=(1-v^2 / c^2)^{-\sfrac{1}{2}}$.
So the condition for pair creation in terms of spacecraft speed is:
\begin{equation}
\gamma \left( 1+\frac{v}{c} \right) \geqslant 2.47 \times 10^{8} 
\end{equation}
corresponding to a $\gamma$ of about $\gamma \approx 1.24 \times 10^{8}$,
or $v/c \approx 1- 3.3 \times 10^{-17}$.

How much energy will be dissipated at this speed or above due to pair production?
There are 412 cosmic microwave photons in the Universe per cubic centimeter
(see Eq.\,(13) below).
Assuming the spacecraft has a crosssectional
area of $A$ (measured in cm$^2$), it will be colliding with approximately
$1.2 \times 10^{13} A$ microwave
background photons per second.
Each created electron-positron pair dissipates $1.6 \times 10^{-13}$ Joules of
energy, thus the viscous-dissipative effect of the cosmic background
on the spacecraft would be
energy dissipation of approximately
$2A$ Joules per second. Assuming an effective cross-sectional area
of say 100 square meters, $A \approx 10^6 \mbox{cm}^2$,
the dissipative effect is about 2 million Joules per second,
but with the time rate of change measured in the rest frame of the cosmic microwave background. In
the spacecraft's rest frame, 1 ``cosmic background second" corresponds to $1/\gamma$
proper time seconds; so energy dissipation is a factor of
$\gamma$ higher, or about $2.5 \times 10^{14}$ Joules per proper time second, which is
a large drag to be overcome by the spacecraft
engines just to keep it moving at constant velocity.
Therefore shutting down the propulsion system when the desired speed is reached
is not an option at large $\gamma$ values. In general,
the drag on the spacecraft scales
as $\gamma^2$, so, for example, at $\gamma \leqslant 10$ the drag will be less than
a few Joules per second and thus negligible.

At the next level, we can imagine technology so advanced that the spacecraft ``hull"
does not contain any ordinary baryonic matter but some other
kind of physical matter field. However,
regardless of how advanced the spacecraft hull material can be,
at sufficiently relativistic velocities pair production of CMB photons via the Schwinger
breakdown of vacuum will become significant.
The Schwinger limit corresponds to the electric field strength at which spontaneous
electron-positron pair
production from the electromagnetic field (photons) becomes energetically favorable:
\begin{equation}
e E_S \frac{h}{m_e c} = 2 m_e c^2
\end{equation}
or
\begin{equation}
E_S = \frac{{m_e}^2 c^3}{\pi e \hbar} \approx 1.4 \times 10^{14} \; ( \mbox{erg/cm$^3$} )^{\sfrac{1}{2}}
\end{equation}
Here $h/ m_e c$ is the electron Compton wavelength, and $e$ is the electronic charge.
Let $\delta$ be the effective “skin depth”
for the interaction/penetration of high energy photons into the ``material" making up
the spacecraft (or its forward shield). Then the colliding gamma
rays will pair-create provided the electric field $E$ associated with their
localization at the length scale $\delta$
and with their frequency $\omega$
in the rest frame of the spacecraft exceeds the Schwinger limit:
\begin{equation}
E \approx \sqrt{\frac{8 \pi \hbar \omega}{\delta^3}} \geqslant E_S = \frac{{m_e}^2 c^3}{\pi e \hbar}
\end{equation}
In terms of the frequency $\omega$ in the rest frame of the spacecraft, the condition
for pair creation is
\begin{equation}
\omega \geqslant \frac{{m_e\!}^4 \, c^6 \delta^3}{8 \pi^3 e^2 \hbar^3} 
= \frac{{E_S}^2}{8 \pi \hbar} \, \delta^3 \approx 7.8 \times 10^{51} \mbox{Hz} \times
[\delta(\mbox{cm})]^3 
\end{equation}
Again with the central CMB frequency of 160GHz,
the blueshift factor to transform the
CMB photon to the rest frame of the spacecraft
is $\omega=\gamma (1+v/c) \times 10^{12}\, \mbox{Hz}$ as before.
So the condition for nucleus-mediated pair creation  in terms of spacecraft speed and
$\delta$ is:
\begin{equation}
\gamma \left( 1+\frac{v}{c} \right) \geqslant 7.8 \times 10^{39} [\delta(\mbox{cm})]^3
\end{equation}
Let us assume the skin-depth $\delta$
for high frequency radiation is on the order of a
micron: $\delta = 10^{-4}\, \mbox{cm}$. This gives
\begin{equation}
\gamma \left( 1+\frac{v}{c} \right) \geqslant 7.8 \times 10^{27}
\end{equation}
corresponding to a $\gamma$ of about $\gamma \approx 4 \times 10^{27}$,
or $v/c \approx 1- 3 \times 10^{-56}$.

Even if the speed of spacecraft is below the limits required for pair production,
simply scattering the microwave
photons creates drag. How big is that drag? To calculate this in order of magnitude, assume each
photon is scattered by reflecting directly back as it hits the spacecraft. This corresponds to a
momentum transfer to each photon of about
\begin{equation}
2 \frac{\hbar \omega}{c} = \gamma \left( 1 + \frac{v}{c} \right)
\, 6.6 \times 10^{-26} \; \mbox{g cm/sec}
= \gamma \left( 1 + \frac{v}{c} \right)
\, 6.6 \times 10^{-31} \; \mbox{Nt sec}
\end{equation}
(as measured in the rest frame of the spacecraft). Taking into account all photons
scattering off the
frontal cross section per unit (spacecraft) proper time,
we obtain a viscous drag force on the spacecraft
of about
\begin{equation}
A (\mbox{cm}^2 )\gamma^2 \left( 1 + \frac{v}{c} \right) \; 8 \times 10^{-18} \mbox{Nt}
\end{equation}
This will become a significant obstacle at speeds for which $\gamma \geqslant 10^{8}$
or $v/c \gtrapprox 1-5 \times 10^{-17}$, which is just below the velocity threshold
for nucleus-mediated pair production.

We summarize various effects caused by interactions with matter and
CMB photons in the table below:

\begin{center}
    \begin{tabular}{| c | c | p{13cm} |}
    $\gamma$ & Speed & $\;\;\;\;\;\;\;\;\;\;\;\;\;\;\;\;\;\;\;\;
    \;\;\;\;\;\;\;\;\;\;\;\;\;\;\;\;\;\;\;\;\;\;\;\;\;\;\;\;\;\;\;\;\;\;\;\;\;\;\;\;$Effect \\ \hline
    1.3 & .69$c$ & pion production in nucleon-nucleon collisions \\ \hline
    9.1 & .994$c$ & anti-proton production in nucleon-nucleon collisions \\ \hline
    $10^8$ & $(1-5 \times 10^{-17})c$ & nucleus-mediated pair
    production in collisions with CMB photons \\ \hline    
    $4 \times 10^{27}$ & $(1-3 \times 10^{-56})c$ & pair production in collisions with CMB photons via
    Schwinger vacuum breakdown \\ \hline       
    \end{tabular}
\end{center}

~

{\noindent \bf Lorentz-transformed Appearance of CMB}

We now turn to examine how the CMB photon distribution would appear to an observer
in the rest frame of the spacecraft. In the rest frame of the CMB, the
photon distribution is thermal
with the Planck density
\begin{equation}
n(\vec{k}) \, d^3 k = \frac{1}{4 \pi^3}\frac{1}{e^{\hbar \omega / k_B T} -1} \, d^3 k 
\end{equation}
where $\omega = c |\vec{k}|$, $k_B$ is Boltzmann's constant, and the left hand side
denotes the number of photons per unit volume with wave vectors lying
in a cell of size $d^3 k$ centered around $\vec{k}$. From Eq.\,(11) we can deduce,
among other things, that the number of photons per unit volume with frequencies
between $\omega$ and $\omega + d \omega$ is
\begin{equation}
n(\omega) \, d \omega = \frac{1}{\pi^2 c^3}\frac{1}{e^{\hbar \omega / k_B T} -1} \, \omega^2 \, d \omega 
\end{equation}
Equation (12) is the more familiar form of the Planck (Bose-Einstein) distribution giving
rise to, for example, the total number of photons per unit volume
\begin{equation}
\int_0^\infty n(\omega )\, d\omega = \frac{2 \, {k_B\!}^3 \, T^3}{\pi^2 c^3 \hbar^3} \, \zeta (3)
\approx 412 \; \mbox{photons/cm}^3
\end{equation}
where $T \approx 2.73 \, ^{\circ}\! K$ for the CMB, and $\zeta$ is the Riemann zeta function.

Consider now a CMB photon with 4-momentum vector
\begin{equation}
p = \hbar \omega (\partial_t + n_x \partial_x + n_y \partial_y + n_z \partial_z )
\end{equation}
in the rest frame of the CMB spanned by the orthonormal basis $(\partial_t ,\partial_x ,
\partial_y , \partial_z )$. We will use units in which $c=1$ unless we explicitly
switch back to standard units for physical interpretation, so here $\vec{n}$ is a unit
3-vector in the direction of the photon's propagation.
The rest frame of the spacecraft is spanned by the basis vectors
\begin{eqnarray}
e_0 & = & \gamma (\partial_t + v \partial_x )  \nonumber \\
e_1 & = & \gamma (v \partial_t + \partial_x ) \nonumber \\
e_2 & = & \partial_y \nonumber \\
e_3 & = & \partial_z
\end{eqnarray}
The same photon is seen in the rest frame of the spacecraft as having
the 4-momentum
\begin{equation}
p = \hbar \omega \gamma (1 - v n_x)\left[
e_0 + \frac{n_x - v}{1- v n_x}\, e_1
+ \frac{n_y}{\gamma (1- v n_x )}\, e_2
+ \frac{n_z}{\gamma (1- v n_x )}\, e_3 \right]
\end{equation}
which is the 4-vector Eq.\,(14) acted on by the Lorentz transformation matrix
\begin{equation}
\Lambda = 
\begin{bmatrix}
\gamma & -\gamma v & 0 & 0 \\
-\gamma v & \gamma & 0 & 0 \\
0 & 0 & 1 & 0 \\
0 & 0 & 0 & 1
\end{bmatrix}
\end{equation}
Similarly,
a photon with 4-momentum vector
\begin{equation}
p = \hbar \omega (e_0 + m_x e_1 + m_y e_2 + m_z e_3 )
\end{equation}
in the rest frame of the spacecraft is seen as having
4-momentum
\begin{equation}
p = \hbar \omega \gamma (1 + v m_x)\left[
\partial_t + \frac{m_x + v}{1 + v m_x}\, \partial_x
+ \frac{m_y}{\gamma (1 + v m_x )}\, \partial_y
+ \frac{m_z}{\gamma (1 + v m_x )}\, \partial_z \right]
\end{equation}
in the rest frame of the CMB,
which is just the 4-vector Eq.\,(18) acted on by the inverse of $\Lambda$
\begin{equation}
\Lambda^{-1} = 
\begin{bmatrix}
\gamma & \gamma v & 0 & 0 \\
\gamma v & \gamma & 0 & 0 \\
0 & 0 & 1 & 0 \\
0 & 0 & 0 & 1
\end{bmatrix}
\end{equation}
The distribution function for photon density transforms
under a coordinate change according to the general
identity
\begin{equation}
\hat{n}(\vec{k}') = n[ \vec{k} (\vec{k}' ) ] \left| \frac{\partial \vec{k} }{\partial \vec{k}'} \right|
\end{equation} 
where $\hat{n}$ is the density in the Lorentz transformed frame, and the right hand side
involves the coordinates $\{ k_x, k_y , k_z \}$ expressed as functions of the transformed
coordinates $\{ k_x ', k_y ', k_z '\}$ where the last term is the Jacobian determinant.
According to Eq.\,(19), we have
\begin{equation}
\vec{k}(\vec{k}') = 
\left[ \gamma(k_x ' + v \omega '), k_y ' , k_z '  \right]
\end{equation}
and
\begin{equation}
\omega^2 = k_x^2 + k_y^2 + k_z^2 = \gamma^2 ( \omega' + v k_x ')^2
\end{equation}
The Jacobian determinant is easily computed:
\begin{equation}
\left| \frac{\partial \vec{k} }{\partial \vec{k}'} \right|
= \gamma \left( 1 + \frac{v k_x '}{\omega '} \right)
\end{equation}
\begin{figure}[htp]
\vspace*{-1in}
\hspace{1.6in}
\centerline{
\input epsf
\setlength{\epsfxsize}{6.100in}
\setlength{\epsfysize}{6.100in}
\epsffile{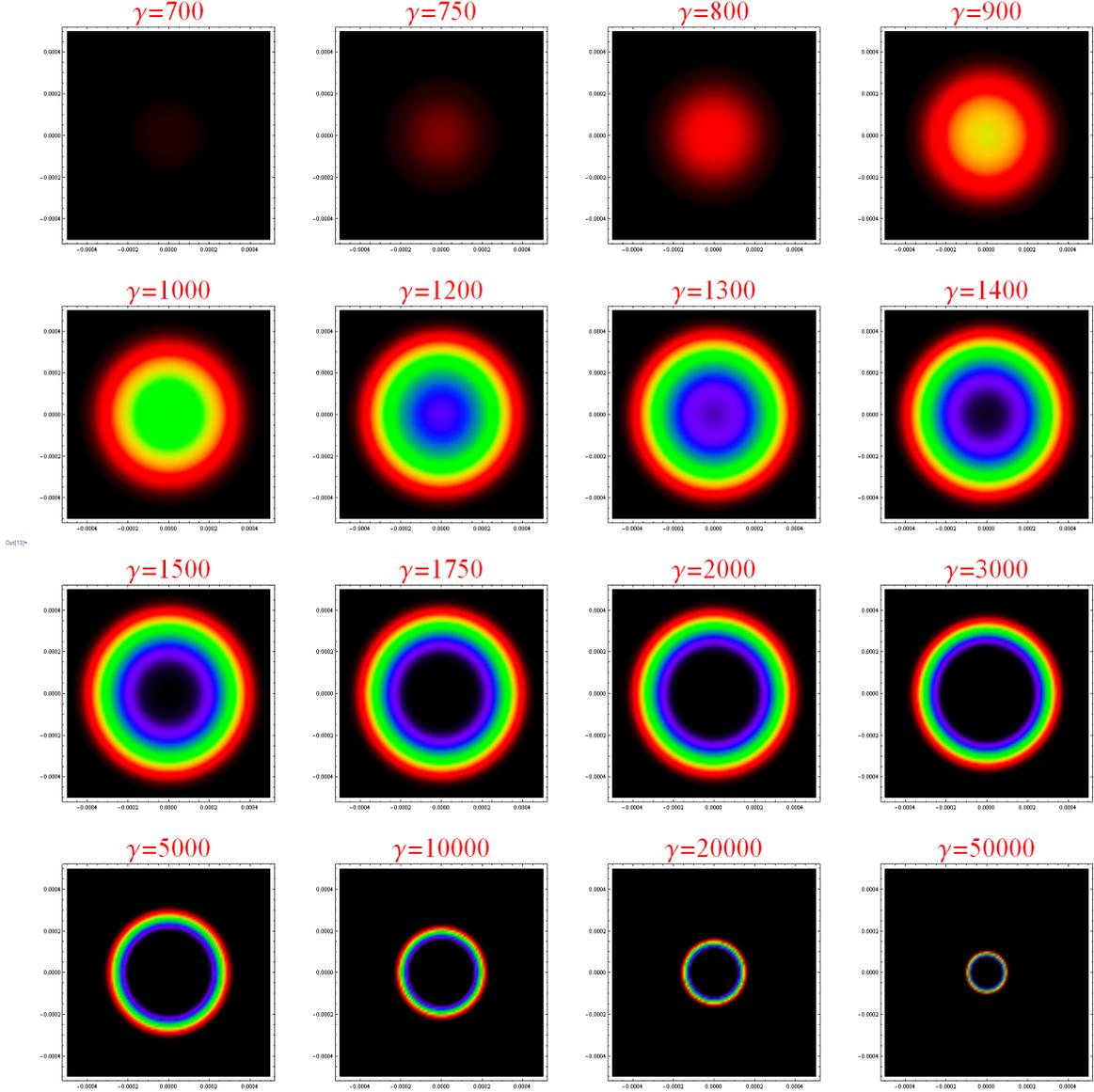}
}
\caption[figure]
{\label{fig:figure1}
How the cosmic microwave background would appear
to a forward
facing observer located in a relativistic spacecraft.
The plots are for increasing spacecraft
speeds (relative to the CMB) as measured by the Lorentz factor $\gamma$.
The appearance is approximately the same as that of a thermal distribution
with latitudinally-varying temperature.
If we use Wien's displacement law, the apparent color of the CMB is
the wavelength at which the spectral radiance of the Planck distribution
reaches its peak. The visible colors form concentric rings confined
to a narrow cone in the polar angle $\theta$,
with the zenith $\theta =0$ fixed along the forward $x$-direction.
The plots depict the apparent ``celestial sky"
parameterized with the $y,z$ coordinates of a unit sphere whose
north pole lies on the $x$-axis.
The first light from blue-shifted CMB photons is visible
in red at around $\gamma = 700$. For an explanation of
the variation in shape and size of the color rings
with increasing $\gamma$ see the discussion following Eq.\,(27).}
\vspace{0.20in}
\end{figure}
Combining Eqs.\,(21)\textendash(24) with Eq.\,(11) and dropping the primes on transformed
quantities for notational convenience, we obtain the CMB photon distribution as seen in
the rest frame of the spacecraft
\begin{equation}
\hat{n}(\vec{k})  =
\frac{1}{4 \pi^3}\frac{\gamma \left( 1 + \frac{v k_x}{\omega} \right)}
{e^{\hbar \gamma (\omega + v k_x ) / k_B T} -1} 
\end{equation}
where $\omega =c \sqrt{k_x^2 + k_y^2 + k_z^2}$.

It is an interesting exercise to plot the appearance of the CMB to a forward
facing observer located in the spacecraft (Fig.\,1).
For qualitative description, we can neglect
the Jacobian factor in the numerator of Eq.\,(25), and notice that for photons moving
opposite to the ``look" direction the wave vector
obeys $k_x = - (\omega /c) \cos \theta$, where $\theta$ measures the angle
between the look direction and the forward ($x$-) direction.
Thus to the forward-facing observer the CMB appears to have a thermal distribution
with latitudinally-varying temperature
\begin{equation}
T(\theta ) = \frac{T}{\gamma \left( 1 - \frac{v}{c} \cos \theta \right)}
=T\frac{\sqrt{1-\frac{v^2}{c^2}}}{\left( 1 - \frac{v}{c} \cos \theta \right)}
\end{equation}
If we use Wien's displacement law and approximate the apparent color of the CMB with
the wavelength at which the spectral radiance of the Planck distribution
reaches its peak, then the color visible at polar angle $\theta$ is given by the wavelength
\begin{equation}
\lambda (\theta) = \frac{b}{T(\theta )} = \frac{b}{T}
\frac{\left( 1 - \frac{v}{c} \cos \theta \right)}{\sqrt{1-\frac{v^2}{c^2}}}
= \frac{b}{T} \left( \gamma - \sqrt{\gamma^2 -1} \cos \theta \right)
\end{equation}
where $b$ is Wien's displacement constant equal
to $2.9 \times 10^{-3}  \, ^{\circ}\rm{K} \, \rm{m}$.
As shown in Fig.1, the visible frequencies are confined to a narrow
cone in $\theta$ around the forward direction, and the constant
color temperatures form rings around $\theta =0$ whose shape and size
vary with increasing $\gamma$. This is closely
related to the phenomeneon of (relativistic) aberration, which
describes how the appearance of an optical source varies
in a geometric, frequency-independent way under a Lorentz transformation~\cite{aberr}.
To understand this variation better, it may help
to examine the angle $\theta$ at which the relativistic Doppler shift of the
incoming CMB photon is zero (i.e., no shift in frequency). The condition for
this to happen is
\begin{equation}
\gamma \left( 1 - \frac{v}{c} \cos \theta \right) =
\gamma - \sqrt{\gamma^2 -1} \cos \theta = 1
\end{equation}
The solution $\theta_0$ to Eq.\,(28) is
\begin{equation}
\theta_0 = \arccos \left( \frac{\gamma-1}{\sqrt{\gamma^2 -1}} \right)
=  \arccos \left(\sqrt \frac{{\gamma-1}}{{\gamma +1}} \right)
\end{equation}
The CMB photons appear blue-shifted to our forward-facing observer
for angles $\theta$ less than $\theta_0$ and
red-shifted for angles $\theta \geqslant \theta_0$.
For small velocities ($\gamma$ close to 1), $\theta_0$ is near $\pi /2$
as would be expected. However, just as in relativistic aberration,
in the limit of large velocities (large $\gamma$) $\theta_0$ approaches zero:
\begin{equation}
\theta_0 = \sqrt{\frac{2}{\gamma}} - \frac{1}{3 \sqrt{2} \; \gamma^{\sfrac{3}{2}}}
+ O \left( \frac{1}{\gamma^{\sfrac{5}{2}}} \right)
\end{equation}
which explains the progressive narrowing at very
large $\gamma$ values of the visible ring in Fig.\,1.

~

{\noindent \bf Signature of CMB Scattered from a Relativistic Spacecraft}

The possibility of detecting radiation associated with
distant relativistic spacecraft
has been discussed in the literature before~\cite{zubrin,viewing,garcia}.
These discussions mostly focus on detecting radiation from spacecraft
engines or light from nearby stars reflecting off the spacecraft.
Our approach is different in that we do not speculate on possible propulsion
technologies but are interested in how a large relativistic object
would interact with the interstellar/intergalactic medium and mainly with the CMB
radiation.
As a baryonic spacecraft travels at relativistic speeds
it will interact with the CMB through scattering
to cause a frequency shift that could be detectable on Earth with current technology.
 
Now we consider the characteristic signature of this
reflected CMB radiation; specifically, CMB photons reflected by
a surface area element of area $A$ moving with velocity $v$
in the $x$ direction. Let the normal to the surface element $A$ be
described by a unit 3-vector $\hat{s}$
in the spacecraft restframe. Then every photon impacting $A$ is reflected back
according to
\begin{equation}
\vec{k} \longrightarrow \vec{k} - 2 (\vec{k} \cdot \hat{s}) \, \hat{s}
\end{equation}
From a far-field perspective the area element $A$ would appear as a point source.
We set up a coordinate system in the CMB
rest frame where $A$ is at $\vec{x}=\vec{v} t=vt\hat{x}$ and the observation point
is at $\vec{r}$. Notice that the time coordinate
is chosen in such a way that the spacecraft's worldline passes through the coordinate
origin at time $t=0$. The radiation detected at $\vec{r}$ at time $t$ would have
been emitted from the spacecraft at the retarded time $\sigma$ which is the time
at the intersection between the past null cone of the event $(\vec{r},t)$ and the
worldline of the spacecraft. Hence the equation for $\sigma$ is
\begin{equation}
c(t - \sigma ) = \sqrt{(x-v \sigma )^2 +y^2 +z^2}
\end{equation} 
The solution to Eq.\,(32) is
\begin{equation}
\sigma^{{\rm ret}\choose{\rm adv}}
= \gamma^2 \left(
t - \frac{vx}{c^2} \mp \frac{1}{c \gamma} \sqrt{\gamma^2 (x-vt)^2 + y^2 + z^2} \right) 
\end{equation}
where the retarded (advanced) solution represents
intersection with the past (future) null cone
of the observation point $(\vec{r},t)$.
We will only use the retarded solution
\begin{equation}
\sigma = \sigma(\vec{r},t)
\equiv \sigma^{({\rm ret})}(\vec{r},t)
= \gamma^2 \left(
t - \frac{vx}{c^2} - \frac{1}{c \gamma} \sqrt{\gamma^2 (x-vt)^2 + y^2 + z^2} \right) 
\end{equation}

Only reflected photons with wave vectors
parallel to $(\vec{r}-\vec{v} \sigma)$
will be detected. There are two Lorentz transformations involved:
First to find the photon distribution in the spacecraft
rest frame according to Eq.\,(25), and second
to transform the distribution reflected according to Eq.\,(31)
back into the CMB rest frame, which is accomplished
by the same formulas as in Eqs.\,(16)--(25).
According to Eqs.\,(16)--(25)
and Eq.\,(31), the spectral intensity of the reflected
radiation in the CMB rest frame
at position $\vec{r}$ and at time $t$ would be
(measured in number of photons of frequency $\omega$ per unit area per unit time
per unit frequency)
\begin{equation}
I (\vec{r},t,\omega ) = \frac{c A}{4 \pi | \vec{r} - \vec{v} \sigma |^2} 
\frac{\omega^2}{\pi^2 c^3}
\frac{\gamma^2 \left[ 1 + \left( \frac{\vec{v}}{c | \vec{r}-\vec{v} \sigma |} \cdot
(\vec{r} - \vec{v} \sigma  - 2 [(\vec{r} - \vec{v} \sigma )\cdot \hat{s}] \hat{s}
\right) \right]^2}
{\exp \left[ \frac{\gamma^2 \hbar \omega}{k_B T} 
\left[ 1 + \left( \frac{\vec{v}}{c | \vec{r}-\vec{v} \sigma |} \cdot
(\vec{r} - \vec{v} \sigma  - 2 [(\vec{r} - \vec{v} \sigma )\cdot \hat{s}] \hat{s}
\right) \right]^2 \right]
-1}
\end{equation}
where $\sigma$ is given by Eq.\,(34) and $\vec{v}=v \hat{x}$.
For a general reflecting
surface $S$ made up of many
small surface elements, the total intensity can be computed as a surface
integral of Eq.\,(35)
\begin{equation}
I_T (\vec{r},t,\omega ) = \frac{\omega^2}{4 \pi^3 c^2}
\oint_{S}
\frac{1}{ | \vec{r} - \vec{v} \sigma |^2} 
\frac{\gamma^2 \left[ 1 + \left( \frac{\vec{v}}{c | \vec{r}-\vec{v} \sigma |} \cdot
(\vec{r} - \vec{v} \sigma  - 2 [(\vec{r} - \vec{v} \sigma )\cdot \hat{s}] \hat{s}
\right) \right]^2}
{\exp \left[ \frac{\gamma^2 \hbar \omega}{k_B T} 
\left[ 1 + \left( \frac{\vec{v}}{c | \vec{r}-\vec{v} \sigma |} \cdot
(\vec{r} - \vec{v} \sigma  - 2 [(\vec{r} - \vec{v} \sigma )\cdot \hat{s}] \hat{s}
\right) \right]^2 \right]
-1}\; dA
\end{equation}
We will limit our attention in this paper
to reflection from a single surface element $A$ as given in Eq.\,(35).
Using Wien's displacement law, the peak frequency of the reflected radiation Eq.\,(35)
is given by
\begin{equation}
\nu_{\rm max} (\vec{r},t)
=\frac{c}{b \gamma^2} \frac{T}{\left[ 1 + \left( \frac{\vec{v}}{c | \vec{r}-\vec{v} \sigma |} \cdot
(\vec{r} - \vec{v} \sigma  - 2 [(\vec{r} - \vec{v} \sigma )\cdot \hat{s}] \hat{s}
\right) \right]^2}
\end{equation}
corresponding to a Planck distribution with effective temperature
\begin{equation}
T_{\rm eff} (\vec{r},t)
=\frac{T}{\gamma^2 \left[ 1 + \left( \frac{\vec{v}}{c | \vec{r}-\vec{v} \sigma |} \cdot
(\vec{r} - \vec{v} \sigma  - 2 [(\vec{r} - \vec{v} \sigma )\cdot \hat{s}] \hat{s}
\right) \right]^2}
\end{equation}
On the other hand, the relative intensity profile is determined by the dimensionless
multiplicative factor
\begin{equation}
i (\vec{r},t,\omega ) = \frac{A}{| \vec{r} - \vec{v} \sigma |^2} 
\gamma^2 \left[ 1 + \left( \frac{\vec{v}}{c | \vec{r}-\vec{v} \sigma |} \cdot
(\vec{r} - \vec{v} \sigma  - 2 [(\vec{r} - \vec{v} \sigma )\cdot \hat{s}] \hat{s}
\right) \right]^2
\end{equation}
\begin{figure}[htp]
\vspace*{-0.3in}
\hspace{1.6in}
\centerline{
\input epsf
\setlength{\epsfxsize}{7.100in}
\setlength{\epsfysize}{4.100in}
\epsffile{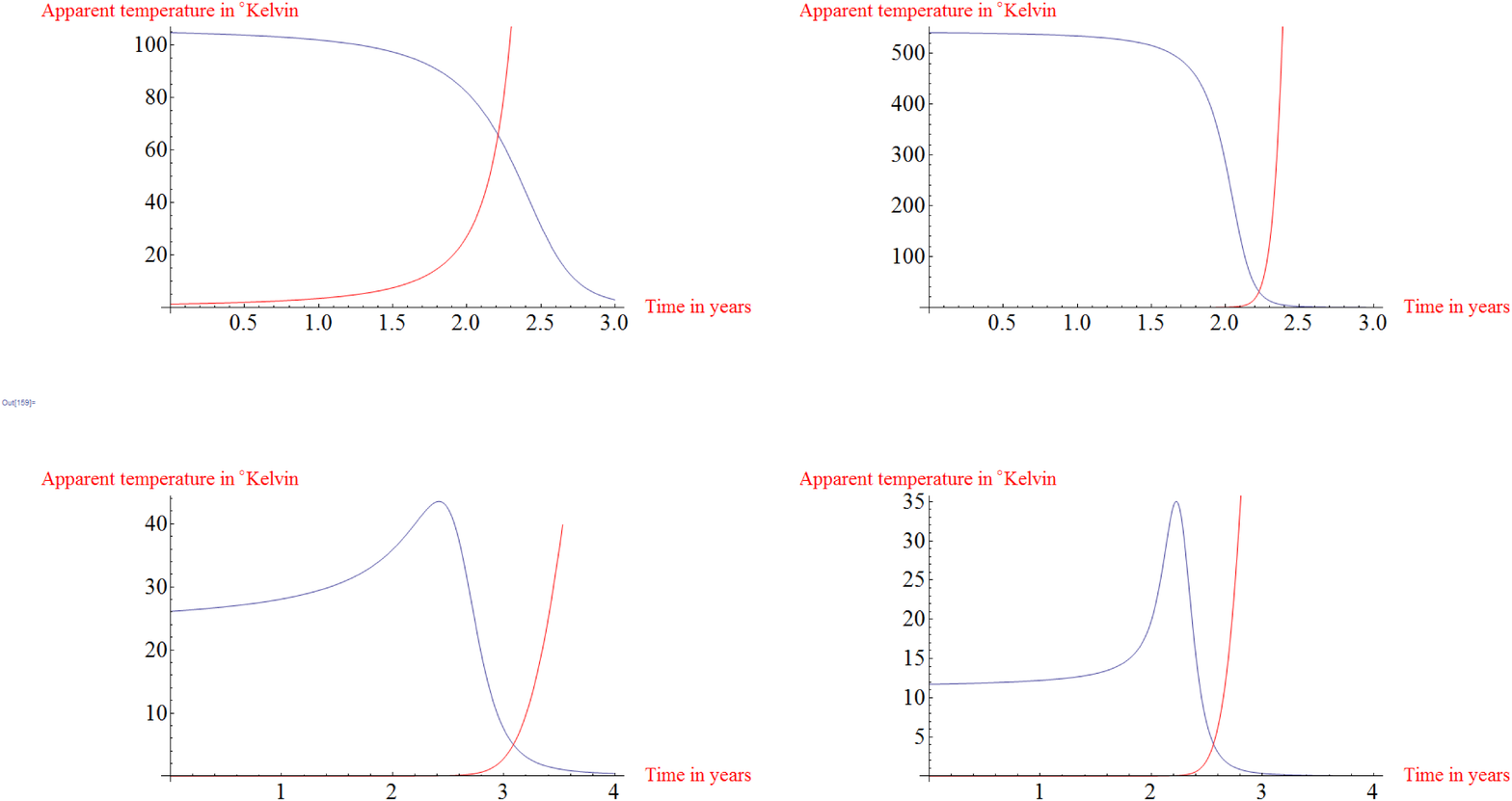}
}
\caption[figure]
{\label{fig:figure2}
Some examples of signal signatures of CMB scattering from relativistic spacecraft.
In all plots the observation point is located at $x= 10$ lightyears,
$y=z=1$ lightyear. The time axis origin is at $t=8$ years (spacecraft is moving
along the $x$-direction and is at
the coordinate origin at time $t=0$). In the upper row, we plot the apparent
(effective) temperature of the signal from a surface element
orthogonal to the $x$-direction (the direction of motion) for spacecraft velocities
$v=0.95 \,c$ and $v= 0.99 \, c$, respectively. In the lower row,
the plots are for a surface element tilted $10^{\circ}$ from orthogonal
to the $x$-direction
but still parallel to the $z$-direction, with the same two velocities.
The red curves are plots of the multiplicative factor Eq.\,(39) in arbitrary units,
showing the relative intensity profile of the received radiation as a function
of time. The decrease in temperature with increasing $\gamma$ in the second
row of plots might seem counter-intuitive at first; but it is due to the same
phenomenon of relativistic aberration discussed above [cf.\ Eq.\,(30)]; namely,
the $10^{\circ}$ tilt in the reflecting
surface is enough to put the reflected beam outside the cone of
blue-shift for the $\gamma$ value corresponding to $v= 0.99 \, c$.
}
\vspace{0.20in}
\end{figure}

Some examples of signal signatures resulting from Eqs.\,(38)--(39) are shown
in Figs.\,2 and 3. The salient features of the signal are a rapid drop in temperature accompanied
by a rapid rise in intensity, along with the motion of the source with respect
to a reference frame fixed to distant quasars, which should be observable. No natural
source of THz to infrared radiation in the known Universe is likely to have time
variability of the kind depicted in Figs.\,2--3. Note that increasing velocity
(increasing $\gamma$) manifests itself in a steepening of the cooling curve, along
with an overall increase in temperature for a surface element oriented orthogonal to the
direction of motion.

\begin{figure}[htp]
\vspace*{-0.1in}
\hspace{1.6in}
\centerline{
\input epsf
\setlength{\epsfxsize}{7.100in}
\setlength{\epsfysize}{2.400in}
\epsffile{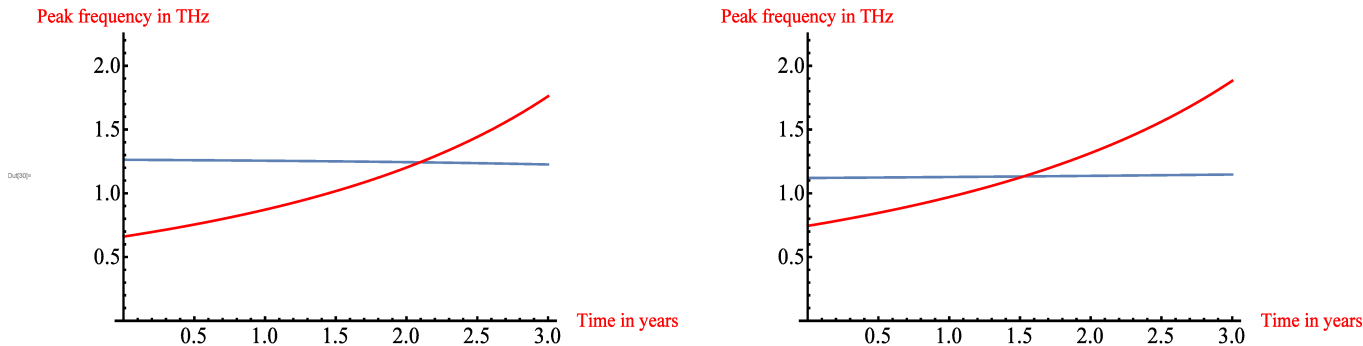}
}
\caption[figure]
{\label{fig:figure3}
A further example of signal signature for
CMB scattering from relativistic spacecraft.
The spacecraft speed in this case is a relatively small
$v=0.64 \,c$ ($\gamma= 1.3$).
As in Fig.\,2, the observation point is located at $x= 10$ lightyears,
$y=z=1$ lightyear. The time axis origin is at $t=8$ years (spacecraft is moving
along the $x$-direction and is at
the coordinate origin at time $t=0$). This time we plot the peak frequency
of the signal; on the left from a surface element
orthogonal to the $x$-direction, and on the right from a
surface element tilted $10^{\circ}$ from that orthogonal
but still parallel to the $z$-direction.
The red curve is again a plot of the multiplicative factor Eq.\,(39) in arbitrary units,
showing the relative intensity profile of the received radiation as a function
of time. The salient features recognizable in Fig.\,2 are much less prominent,
owing to the comparatively small value of $\gamma$ in this case.
}
\vspace{0.20in}
\end{figure}

Finally, we note the similarity of
CMB scattering from macroscopic relativistic
objects to the Sunyaev-Zeldovich (S-Z) effect in cosmology.
The S-Z effect is the spectral distortion
of the CMB caused by inverse Compton scattering of microwave
background photons from energetic intracluster thermal electrons
(intracluster referring to clusters of
galaxies)~\cite{SZ1,SZ2}.
S-Z scattering has been observed~\cite{birkinson,carlstrom}.
The intracluster electrons have temperatures of order
3 to 15 keV, which correspond to
gamma factors of $\gamma \approx 1.01 - 1.03$, which correspond to
thermal velocities of $ v / c \approx 0.15 - 0.25$. Therefore,
while these electrons are relativistic,
they are far from the ultra-relativistic speeds we are interested in.
However, the observed S-Z effect in cosmology
is qualitatively similar to what we discuss above.

On the other hand, merger shocks in the collisions
of clusters of galaxies can produce non-thermal
electrons at energies from 300MeV ($\gamma \approx 300$)
up to $10^6$MeV ($\gamma \approx 2 \times 10^6$),
so these electrons are
more relevant to our discussion~\cite{mergershocks}.
In fact, these ultra-relativistic
electrons, protons, and heavier ions
accelerated by merger shocks loose
energy via inverse Compton, pair production, and
other effects mediated by collisions with the cosmic microwave photons,
much as we discussed for the ultrarelativistic
spaceship example. The S-Z effect in this case,
which is exactly the same as the spaceship
example except the relativistic object
is a single particle, should be similar provided it is
observable.

~

{\bf \noindent Summary}

We have explored the physical interaction of a relativistic spacecraft
with the interstellar and intergalactic mediums. Our main discussion
focused on the interaction with the CMB where very little information
currently exists and seems to have been overlooked. Central to our discussion
was not how to obtain relativistic speeds but the consequences of traveling
that fast. We discussed the interaction with baryonic matter in terms
of high speed collisions to present a complete picture. In general
one can imagine the same interactions that occur in a particle accelerator
to occur between relativistic spacecraft and interstellar matter.
Our assumption that matter-matter interactions can be dealt with when
civilization can build relativistic spacecraft may prove false
and may be a barrier that will prevent space travel with a large $\gamma$.

We looked at two special reference frames, the spacecraft frame,
and the rest frame of the CMB to understand how the CMB is distorted
or acquires an aberration from those view points.
The scattering of the CMB from a relativistic spacecraft is
very similar to the Sunyaev-Zedovich effect where the inverse
Compton scattering instead occurs at a macroscopic level.
Our calculation for what an observer on earth could detect predicts
a very unusual signature that is unlikely to be caused by
any naturally occurring object
in the known universe.
This result is independent of propulsion technology, but the ability
to detect the signal from Earth depends on available detector
technologies. We are currently working
to predict how far can we see this signature given our current capability.

\renewcommand{\thefootnote}
{\ensuremath{\fnsymbol{footnote}}}
\addtocounter{footnote}{2}
~~~

~~~



~~~~

\end{document}